\documentclass[conference]{IEEEtran}

\usepackage{url}
\usepackage{cite}
\usepackage{amsmath,amssymb,amsfonts}
\usepackage{algorithmic}
\usepackage{graphicx}
\usepackage{textcomp}
\usepackage[dvipsnames,table,xcdraw]{xcolor}
\usepackage{siunitx}
\usepackage{booktabs}
\def\BibTeX{{\rm B\kern-.05em{\sc i\kern-.025em b}\kern-.08em
    T\kern-.1667em\lower.7ex\hbox{E}\kern-.125emX}}
\usepackage[export]{adjustbox}
\usepackage[
    bookmarks,
    pdftex,
    colorlinks=true,
    pagebackref=true,
    backref=page,
    linkcolor={red!50!black},
    filecolor={green!50!black},
    citecolor={green!50!black},
    urlcolor={blue!80!black}
]{hyperref}
\usepackage{xcolor}
\usepackage{booktabs}

\graphicspath{{img/}}

\begin{document}

\title{Vaccine Hesitancy on YouTube: \\a Competition between Health and Politics}


\author{\IEEEauthorblockN{Yelena Mejova}
\IEEEauthorblockA{\textit{ISI Foundation} \\
Via Chisola 5, Turin, Italy \\
yelenamejova@acm.org}
\and
\IEEEauthorblockN{Michele Tizzani}
\IEEEauthorblockA{\textit{Department of Applied Mathematics and Computer Science} \\
\textit{Technical University of Denmark}\\
Richard Petersens Plads, 2800, Copenhagen, Denmark \\
micti@dtu.dk}}



\maketitle

\begin{abstract}

YouTube has rapidly emerged as a predominant platform for content consumption, effectively displacing conventional media such as television and news outlets. 
A part of the enormous video stream uploaded to this platform includes health-related content, both from official public health organizations, and from any individual or group that can make an account.
The quality of information available on YouTube is a critical point of public health safety, especially when concerning major interventions, such as vaccination. 
This study differentiates itself from previous efforts of auditing YouTube videos on this topic by conducting a systematic daily collection of posted videos mentioning vaccination for the duration of 3 months. 
We show that the competition for the public's attention is between public health messaging by institutions and individual educators on one side, and commentators on society and politics on the other, the latest contributing the most to the videos expressing stances against vaccination.
Videos opposing vaccination are more likely to mention politicians and publication media such as podcasts, reports, and news analysis, on the other hand, videos in favor are more likely to mention specific diseases or health-related topics.
Finally, we find that, at the time of analysis, only 2.7\% of the videos have been taken down (by the platform or the channel), despite 20.8\% of the collected videos having a vaccination hesitant stance, pointing to a lack of moderation activity for hesitant content.
The availability of high-quality information is essential to improve awareness and compliance with public health interventions.
Our findings help characterize the public discourse around vaccination on one of the largest media platforms, disentangling the role of the different creators and their stances, and as such, they provide important insights for public health communication policy.

\end{abstract}


\section{Introduction}
\label{sec:intro}

A January 2025 report by Nielsen has found that streaming platforms accounted for the largest share of all TV viewing formats, and that YouTube is the top used service in the United States \cite{neilsen25youtube}. 
YouTube has been acknowledged as an important resource for science communication as early as 2010, when it was found that the CDC's YouTube channel contributed about 12\% of engagement around videos about the ongoing H1N1 influenza epidemic \cite{pandey2010youtube}.
But even then, 23\% of the videos on the topic were found to be misleading.
During the COVID-19 pandemic, the platform has announced a blanket ban on anti-vaccine misinformation,\footnote{https://www.businessinsider.com/non-us-anti-vaxxers-are-still-on-youtube-despite-ban-2021-9} but the enforcement remained lax, and recent audits have found a variety of anti-vaccination content \cite{li2022youtube,gruzd2023facebook}. 
This may be understandable, since de-platforming channels promoting vaccination hesitancy would cost YouTube millions in profits \cite{ccdh2021pandemic}.

Although public health professionals and organizations are increasing their presence on social media platforms, they may be competing with potentially millions of other organizations and individuals for the public's attention.\footnote{https://www.straitstimes.com/opinion/apple-cider-vinegar-how-social-media-gave-rise-to-fraudulent-wellness-influencers}
``Influencers'' who garner large followings may have audiences larger than the accounts of legitimate public health institutions, and may wield their power to either educate and entertain the public, but also to possibly promote unproven or harmful health advice.
Such harmful advice may be targeted specifically to already vulnerable populations, such as those having serious illness or those who have had a negative interaction with physicians or unsuccessful treatments, and who have developed a distrust for mainstream medicine.\footnote{https://www.apa.org/topics/journalism-facts/false-health-information}
Since the COVID-19 pandemic, the vaccination debate in particular has been politicized, garnering attention from pundits and commentators having limited expertise in public health or medicine.

In the present study, we shed light on the landscape of the YouTube channels posting content mentioning vaccination by collecting a unique dataset of daily collections of videos over a 3-month period.
We use the latest NLP methods to detect the video's stance on the benefits of vaccination, and examine the channels contributing to the videos supporting or opposing vaccination.
Examining the metadata and transcripts of 3496 videos, we found that about 21\% of these had a stance against vaccination. 
These videos were much more likely to come from channels associated with individuals than organizations, and from those on the topics of politics and society (as opposed to knowledge or health). 
Worryingly, several channels self-identifying as ``doctors'' have posted such videos. 
Further, they are much more likely to mention politicians and media coverage, unlike videos with a stance favoring vaccines, which are more likely to mention specific illnesses which vaccines alleviate. 
Finally, we find that only 2.7\% of the videos against vaccination have been taken down (by the platform or the poster), indicating a lax moderation regime.






\section{Related Work}
\label{sec:relwo}


Exposure to low-quality information on internet, and social media in particular, has long been connected with vaccine hesitancy. 
Respondents to a 2021 survey of 27 EU Member States who favored social media as a source of news were much more likely to be vaccine hesitant \cite{mascherini2022social}, while Twitter users in the U.S. who followed politicized news sources were less likely to vaccinate during COVID-19 \cite{rathje2022social}. 
Some estimates put the effects of this information into perspective: for instance, it is estimated that, in Canada, COVID-19 misinformation has cost at least 2,800 lives and \$300 million in hospital expenses \cite{major2023covid}. 
Thus, estimating the quality of vaccine information online has been an active area of research.



YouTube serves as a significant channel for disseminating information, including health and science topics, offering an accessible alternative to traditional learning \cite{kohler2021potentials}. 
However, the platform's open nature means the validity of health-related content, especially concerning vaccination, can range from reliable to dangerously misleading (see \cite{haslam2019youtube} for a literature review).
Some of the earliest research on YouTube content around vaccination comes from 2007 \cite{keelan2007youtube}, which performed an audit of videos mentioning ``vaccination" or ``immunization''.
The authors found that 32\% of the videos were negative, and 20\% were ambiguous about immunization. 
This trend was confirmed by later studies, highlighting common anti-vaccination arguments including alleged side effects, autism links, pharmaceutical lobbying, and adjuvant concerns \cite{donzelli2018misinformation, lahouati2020spread}. 
The COVID-19 pandemic intensified this, with significant content contradicting established health organizations \cite{li2022youtube}. 
This issue extends beyond YouTube, with anti-vaccine messages spreading across platforms \cite{gruzd2023facebook}, although YouTube's recommendation algorithms may sometimes steer users towards pro-vaccine content \cite{ng2023exploring}.

The source of information critically influences its reception. 
Research distinguishes between professionally generated content (PGC) from entities and content from individual creators \cite{welbourne2016science, kim2012institutionalization}. 
In health communication, signals of authority (e.g., medical credentials, institutional affiliation) are often used \cite{haslam2019youtube}, though anti-vaccination content may mimic scientific authority through specific terminology \cite{yiannakoulias2019expressions}. 
Notably, production quality doesn't always correlate with scientific accuracy or creator credentials \cite{munoz2016typologies}.
In this study, we focus on the signals of authorities the channels may project, as organizations or as individuals.
Note that the claims of accreditation or affiliation may be wholly or partially erroneous. 
However, it is the projection of authority on which we focus in this study.
Also note that the quality of the video may not necessarily correlate with the authoritativeness of its creative.
For instance, a detailed analysis of 190 videos of popular scientific YouTube channels found that non-affiliated channels such as \emph{SciShow} and \emph{Veritasium} may have high production values and garner followers in the hundreds of thousands \cite{munoz2016typologies}.


Because most of the previous literature has used a laborious manual process to assess video content, the labeling of the video's stance on the vaccination or immunization was often done manually \cite{gruzd2023facebook,allgaier2019science,ledwich2019algorithmic,briones2012vaccines,yiannakoulias2019expressions}.
While machine learning methods like SVMs were explored \cite{ng2023exploring}, recent Large Language Models (LLMs) offer a more scalable and efficient approach \cite{fields2024survey, espinosa2024use}, demonstrating effectiveness in classifying vaccine stances on social media \cite{soojong2024accuracy}.
In this work, we use the latest available technology: the Large Language Models (LLMs), which have been shown to perform well on Natural Language Processing (NLP) tasks \cite{fields2024survey}.
In particular, it has been shown that LLMs provide a faster and more efficient option compared to traditional methods for the task of vaccine stance classification \cite{espinosa2024use}, whereas another showed the performance of ChatGPT (\texttt{gpt-3.5-turbo-0613} model) to attain accuracy between 0.72 and 0.88, when tested on Twitter and Facebook posts \cite{soojong2024accuracy}.
In this study, we use the latest available ChatGPT model (\texttt{gpt-4o-mini}) and manually evaluate its output.

\section{Data \& Methods}

\subsection{Data Collection}

To achieve the aim of a temporal sample of YouTube videos mentioning vaccination, we devised a daily collection routine, which we have executed since November 20, 2024. 
The dataset for this study considered all collected data up until February 19, 2025, resulting in 3 months of data. 
The collection is ongoing for future work.
Every day, three actions are taken: search for the previous day's videos, a selection of videos of interest from the search, and crawl additional metadata of the selected videos.

The daily search queries the YouTube Search API\footnote{\url{https://developers.google.com/youtube/v3/docs/search/list}} for videos posted in the previous 24 hours with one or more of the following keywords: \emph{vaccination, vaccine, antivaxx, antivaxxer, novax, provax}, following the same logic as in \cite{crupi_echoes_2022, Paoletti2024-hk} (note that the search performs stemming by default and forms of words such as ``vaccinations" are also retrieved).
The settings are set as follows: region as ``US", language as English, order of videos by relevance.
Up to 500 video IDs are collected.
Using these IDs, their metadata is collected using YouTube Video API call,\footnote{\url{https://developers.google.com/youtube/v3/docs/videos}} as well as the transcript using the YouTube Transcript API library.\footnote{\url{https://pypi.org/project/youtube-transcript-api/}} 
The collected videos are then passed through several filters to select those more likely to be concerning the topic of vaccination, and to have enough metadata for analysis.
Specifically, we select videos that have a language detected as English (passing about 32\% of the videos), that are between 1 and 60 minutes in length (down to 17\% of total), have a transcript (13\%), and have at least one keyword in the title and transcript, or if there are no keywords in the title, have at least 3 keywords in the transcript.
On average, about 9\% of the search results pass these filters and are added to the dataset.
We then collect additional metadata about the videos, including that about the channels that posted them.
In total, 3496 videos were collected.


\subsection{Data Annotation}

\subsubsection*{Vaccination Stance}

To distinguish between videos that may contain vaccine-hesitant content, we use the most recent LLM technology to label the transcript snippets.
Focusing on the content relevant to vaccination, we extract 100 word excerpts from the video transcript around the keywords of interest, and select up to 20 for labeling (excerpts for videos having more than 20 are randomly sampled).
The \num{23820} excerpts (on average, 6.8 per video) were then annotated using the \texttt{gpt-4o-mini} model with the following labels: ``in favor of vaccination", ``against vaccination", ``not in favor or against vaccination".
The prompt used for the annotation can be seen in the Appendix A\ref{apx:prompt}.
Two authors annotated a subset of 50 excerpts, the inter-annotator agreement was at 100\% overlap, or Cohen's Kappa of 1. 
The 84\% of human labels matched with those of the LLM, though only two of the mismatches were against vs.~ in favor, and other mismatches involved the neutral label.
Given this performance of the model, we extend its excerpt labels to the videos. 
To do so, we compute the average score over all excerpts in a video (coding \emph{against} as -1, \emph{in favor} as 1, and \emph{not in favor or against} as 0). 
We then used a Likert-like scoring for the videos using the following ranges: ``strongly in favor" [0.5, 1], ``in favor" (0, 0.5), ``neutral" [0], ``against" (-0.5, 0), and ``strongly against" vaccines [-1.0, -0.5]. 
The distribution of labels and their descriptive statistics can be seen in Table \ref{tab:dataset}.

\begin{table}[t]
\caption{Video categories, score ranges, number of videos and channels in each, and the proportion of videos to a channel.}
\begin{center}
\begin{tabular}{lrrrr}
\toprule
Category & Score range & \# Videos & \# Channels & V/C\\
\midrule
strongly in favor & [0.5, 1]     & 1188 & 880 & 1.35 \\
in favor          & (0, 0.5)     & 744  & 585 & 1.27 \\
neutral           & [0]          & 836  & 718 & 1.16 \\
against           & (-0.5, 0)    & 353  & 285 & 1.24 \\ 
strongly against  & [-1.0, -0.5] & 375  & 291 & 1.29 \\
\bottomrule   
\end{tabular}
\label{tab:dataset}
\end{center}
\end{table}

\subsubsection*{Channel Classification}

To further understand the role of the most active content creators, we also classify the YouTube channels with the most videos for each stance. Specifically, for each stance, we select the first 30 channels with the most videos in our dataset and manually annotate them. This threshold of 30 was chosen as it represents the most active creators before entering the long tail of the distribution, where channels contributed far fewer videos each.
To define the channel classes, we consider two criteria: the \textit{channel owner} and the \textit{channel category}. 
In particular, the channel owner can be either an \textit{individual}, an \textit{organization}, or neither (\textit{other}). 
For the content category we focus on five different classes that are relevant for our study. 
The definition of the category was extrapolated from the primary focus of the channel, as indicated by the thematic content of the videos produced and the content creators' self-declaration of intent. 
The categories are the following: \textit{health} for the channels focusing on health-related topics; \textit{news} for all the channel self-declared news outlets, \textit{opinion} for all the channels not self-declared as official news or information sources, \textit{science} for channels self-declared with a focus on science, \textit{doctor} for channels hosted by self-identified doctors, and \textit{others}.

\section{Results}

\subsection{Video Characterization}

\begin{figure*}
    \centering
    \includegraphics[width=0.8\linewidth]{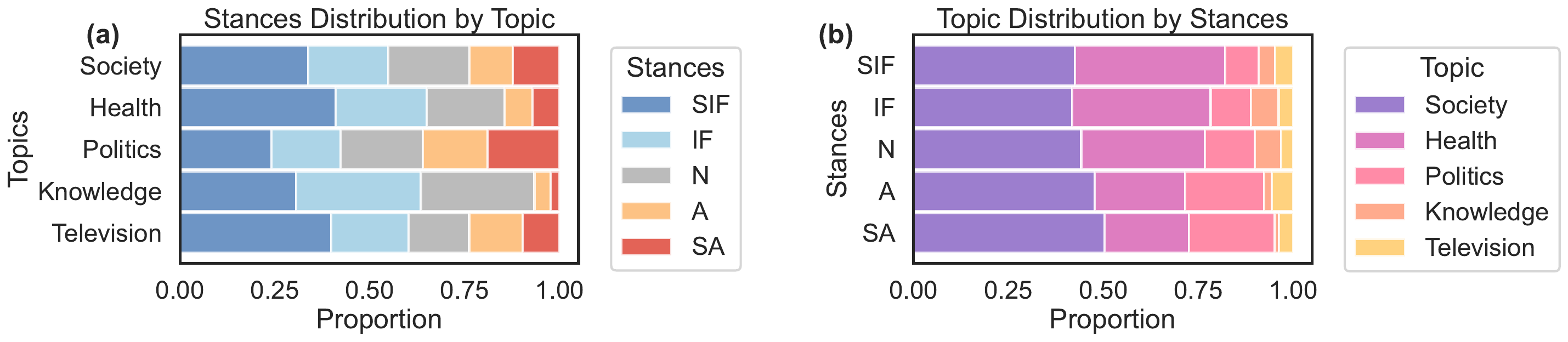}
    \caption{The stacked bar plots show the stance distributions by topic (a) and topic distribution by stances (b). The left plot has topics on the y-axis and proportion on the x-axis. The right plot has stances on the y-axis and proportions on the x-axis.}
    \label{fig:video_stance}
\end{figure*}

We begin by considering the topical categories associated with the collected videos, which were labeled with different stances.
Figures\ref{fig:video_stance} (a-b) show these relationships for the five most frequent topics in our dataset: Health, Knowledge, Politics, Society, and Television. 
The plots present complementary views: the distribution of stances within each topic (a), and the distribution of topics within each stance (b). 
Overall, Society and Health were the most common topics assigned to videos in our collection.
Figure \ref{fig:video_stance}a displays the distribution of stances within each topic. 
Videos categorized under Health predominantly exhibit ``Strongly In Favor" (SIF) or ``In Favor" (IF) stances. 
On the other hand, topics related to Society and Politics show a markedly higher proportion of ``Against" (A) and ``Strongly Against" (SA) stances. The Knowledge and Television topics display more intermediate distributions, generally showing lower proportions of hesitant (A and SA) stances compared to Society and Politics.
Figure \ref{fig:video_stance}b shows the topic distribution within each stance, reinforcing these observations. The SIF and IF stance categories are mainly composed of videos tagged with the Health topic. 
In contrast, the A and SA stance categories draw heavily from videos tagged with Society and Politics.
%

Besides the general categories, the videos are also annotated with ``tags" which are not standardized.
We lower-case these and remove all non-alphanumeric characters. 
We then compute the odds ratio that a tag appears in a video labeled as SIF and not in SA (low scores of this metric also tell us the reverse).
Table \ref{tab:toptags} shows the top tags for the two labels. 
Those associated with the videos labeled as SIF include numerous mentions of diseases and conditions which vaccines prevent, such as \emph{hpv}, \emph{cervical cancer}, \emph{measles}, \emph{human papilloma virus}, \emph{whooping cough}, and \emph{hiv}.
It also specifically mentions \emph{women's health} and developing regions such as \emph{africa} and \emph{india}.
The US \emph{Centers for Disease Control and Prevention} and \emph{Department of Health \& Human Services (HHS)} are also prominent.
On the other hand, the tags more likely to appear in the SA videos mention politics (\emph{republicans}, \emph{RFK} in relation to Robert F. Kennedy Jr. who was nominated by the US president as the Secretary of Health and Human Services, and \emph{Elizabeth Warren}, a US senator, as well as \emph{Trump} and \emph{Biden}, both US presidents). 
A political slogan is also present: \emph{make america healthy again}, echoing the catchphrase made popular by Donald Trump (\emph{Make America Great Again}).
Note that no public health institution is mentioned in these, but instead are \emph{reports}, \emph{newsanalysis} and \emph{podcast}. 

\begin{figure*}[ht]
    \centering
    \includegraphics[width=1\linewidth]{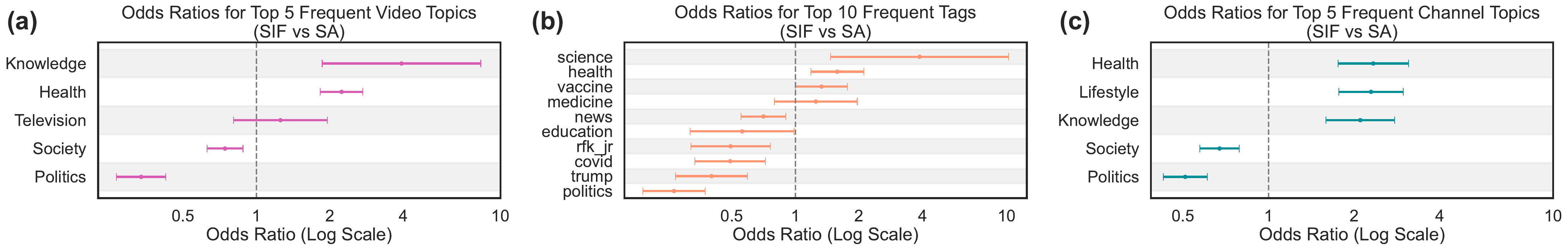}
    \caption{Odds Ratios (ORs) with 95\% Confidence Intervals (CIs), comparing features associated with videos classified as Strongly In Favor (SIF) versus Strongly Against (SA) vaccination, with SA as the reference group. An OR $>$ 1: feature is more associated with the SIF stance, OR $<$ 1: with the SA stance. 
    The panels show results for: (a) the top 5 frequent video topics, (b) the top 10 frequent grouped video tags, and (c) the top 5 frequent channel topics.}
    \label{fig:odds_ratio_comparison} 
\end{figure*}

Next, we stem the tags, group them, and compute the odds ratios (OR) of the tag group appearing in videos labeled as SIF against those labeled SA.
Figure \ref{fig:odds_ratio_comparison}b shows the OR for the top 10 most frequent grouped tags. 
We find clear thematic distinctions: tags related to ``science" and ``health" are significantly more likely to appear in SIF videos, with odds ratios substantially greater than 1 and confidence intervals entirely above the null value. 
In contrast, tags associated with ``news", ``education", ``rfk\_jk", ``covid", ``trump", and ``politics" all show odds ratios significantly less than 1, indicating these terms were disproportionately more common in SA videos. 
Tags directly mentioning ``vaccine" and ``medicine" did not show a statistically significant difference between the groups. 

\begin{table*}[t]
\caption{Top tags of videos by odds ratio (shown in parentheses): on the left, the tags that are more likely to appear in videos labeled as ``strongly in favor'', on the right, those more likely to appear in videos that are ``strongly against" vaccines.}
\label{tab:toptags}
\begin{center}
\begin{tabular}{p{8cm}p{0.2cm}p{8cm}}
\toprule
\textbf{Strongly in Favor} & & \textbf{Strongly Against} \\
\midrule
local (0.02), cervicalcancer (0.04), hpv (0.04), womenshealth (0.05), measles (0.06), centersfordiseasecontrolandprevention (0.06), rsv (0.06), humanpapillomavirus (0.07), vaccinedevelopment (0.07), whoopingcough (0.07), currentaffairs (0.08), africa (0.08), howvaccineswork (0.08), cbs (0.09), hivvaccine (0.09), vaccinescience (0.09), hiv (0.09), smallpox (0.09), india (0.09), hhs (0.09) & &
reports (5.05), disease (4.34), uk (4.23), newsanalysis (3.45), covidvaccine (3.10), republicans (2.92), biology (2.72), podcast (2.64), rfk (2.60), antivaccine (2.39), elizabethwarren (2.39), makeamericahealthyagain (2.39), businessnews (2.39), republican (2.27), trump (1.86), medicine (1.86), biden (1.86), election (1.86), economy (1.86), senateconfirmationhearing (1.86) \\

\bottomrule
\end{tabular}
\end{center}
\end{table*}

\subsection{Channel Characterization}

Next, we focus on the channels posting these videos. 
We begin by examining the videos with each of the annotated label categories. 
As can be seen in Table \ref{tab:dataset}, the proportion of videos to channels is fairly stable across the categories, at about 1.3 videos per channel. 
Table \ref{tab:channels} shows the top 10 channels in each category, and for the top 3 -- the number of subscribers and an excerpt from their description (full descriptions omitted for brevity).
Note that, since the labels are per video, it is possible for a prolific channel to contribute videos to several label categories.
For instance, ``Forbes Breaking News" is the top contributor of videos labeled as SIF and N.
Similarly, a channel of ``Dr. John Campbell" is in the list of top contributing channels both in the ``against" and ``strongly against" categories.
Furthermore, the channel ``MicrobeTV" contributes videos to both IF and A categories.

In general, we find news organizations in the SIF and IF categories, as well as health-oriented organizations (WHO) and educational efforts (HIV RNA Test Guide). 
On the other hand, more individuals are visible at the top of the ranking for the SA and A categories (Dr. John Campbell, Russell Brand); however, self-described ``news" channels are also present, including ``Rebel News" and ``GBNews''. 

We find that the potential reach of these channels varies widely, with individual-centered channels, such as ``Dr. John Campbell" having viewership in the millions, whereas, for instance, WHO's channel has smaller (but still substantial) 930K subscribers.
Throughout the dataset, we note that there isn't a single channel that dominates the sample: the top contributing channel -- ``HIV RNA Test Guide" -- is responsible for only 0.8\% of videos in the dataset, indicating the breadth of the sampling method used in this study.


\begin{table*}[t]
\caption{For each video category, the top 10 contributing channels (in parentheses, number of contributed videos), and for the top 3, the number of subscribers and an excerpt from the channel description.}
\begin{center}
\begin{tabular}{p{3cm} l p{12cm}}
\toprule

Channel  & Subscribers & Description excerpt \\
\midrule

\multicolumn{3}{c}{\textbf{Strongly in favor of vaccines (SIF)}} \\
Forbes Breaking News (12) & 4.64M & [none] \\
HIV RNA Test Guide (8) & 5.14K & a platform where people can learn about common sexually transmitted diseases i.e STDs. \\
World Health Organization (WHO) (7) & 930K & The official public health information Youtube channel of the World Health Organization (WHO) \\
\multicolumn{3}{l}{Others: Dr Kat, Epidemiologist (7), Citizen TV Kenya (7), TTT Live Online (6), CP24 (6), 9NEWS (6), GoodRx (5), Channels Television (5)} \\

\midrule
\multicolumn{3}{c}{\textbf{In favor of vaccines (IF)}} \\
HIV RNA Test Guide (19) & 5.14K & a platform where people can learn about common sexually transmitted diseases i.e, STDs. \\
Dawah Institute (9) & 3.53K & dawahinstitute.org \\
MicrobeTV (9) & 135K & I'm Vincent Racaniello, Earth's virology Professor and I believe that education should be free. \\
\multicolumn{3}{p{17cm}}{Others: CreakyJoints (7), Citizen TV Kenya (6), Forbes Breaking News (5), Casina Pio IV (5), WION (5), Medscape (5), KGW News (4)} \\

\midrule
\multicolumn{3}{c}{\textbf{Neutral of vaccines (N)}} \\
Forbes Breaking News (7) & 4.64M & [none]  \\
Vejon Health (5) & 157K & At Vejon Health, we want to make the world a better place for you and future generations by improving the understanding and management of complex diseases.  \\
Stuff (4) & 68K & Stuff is New Zealand's largest and most popular news site. \\
\multicolumn{3}{p{17cm}}{Others: Dr Neha Taneja's Community Medicine (4), WION (4), NTV Kenya (4), HIV AIDS PREVENTION (4), Chuck Miller Plays (4), KBC Channel 1 (3), Kenya Digital News (3)} \\

\midrule
\multicolumn{3}{c}{\textbf{Against of vaccines (A)}} \\
Vejon Health (9) & 157K & At Vejon Health, we want to make the world a better place for you and future generations by improving the understanding and management of complex diseases.  \\
Dr. John Campbell (8) & 3.19M & My name is John Campbell and I am a retired Nurse Teacher and former clinical nurse based in England. [...] Disclaimer; These media including videos, book, e-books, articles, podcasts are not peer-reviewed. \\
Russell Brand (8) & 6.84M & Something ancient and original is Awakening within us. Here we can learn from elders and grow together in Good Faith, flawed but rising together. \\
\multicolumn{3}{p{17cm}}{Others: The Jimmy Dore Show (5), GBNews (4), Sara Gonzales Unfiltered (4), TalkTV (4), Rebel News (4), The Next News Network (3), MicrobeTV (3)} \\

\midrule
\multicolumn{3}{c}{\textbf{Strongly against of vaccines (SA)}}\\
Find the Best (12) & 2.77K & At Find the Best, we are dedicated to helping you make informed decisions by providing comprehensive information on equipment, companies, products, and services. \\
GBNews (9) & 1.75M & Keep up to date with GB News at gbnews.com or on X \@GBNEWS \\
Dr. John Campbell (9) & 3.19M & My name is John Campbell and I am a retired Nurse Teacher and former clinical nurse based in England. [...] Disclaimer; These media including videos, books, e-books, articles, and podcasts, are not peer-reviewed. \\
\multicolumn{3}{p{17cm}}{Others: The Jimmy Dore Show (8), Redacted (7), Russell Brand (6), Kim Iversen (5), Rebel News (5), Viva Frei (3), Starkey Farmstead (3)} \\

\bottomrule
\end{tabular}
\label{tab:channels}
\end{center}
\end{table*}

\begin{figure*}
    \centering
    \includegraphics[width=0.92\textwidth]{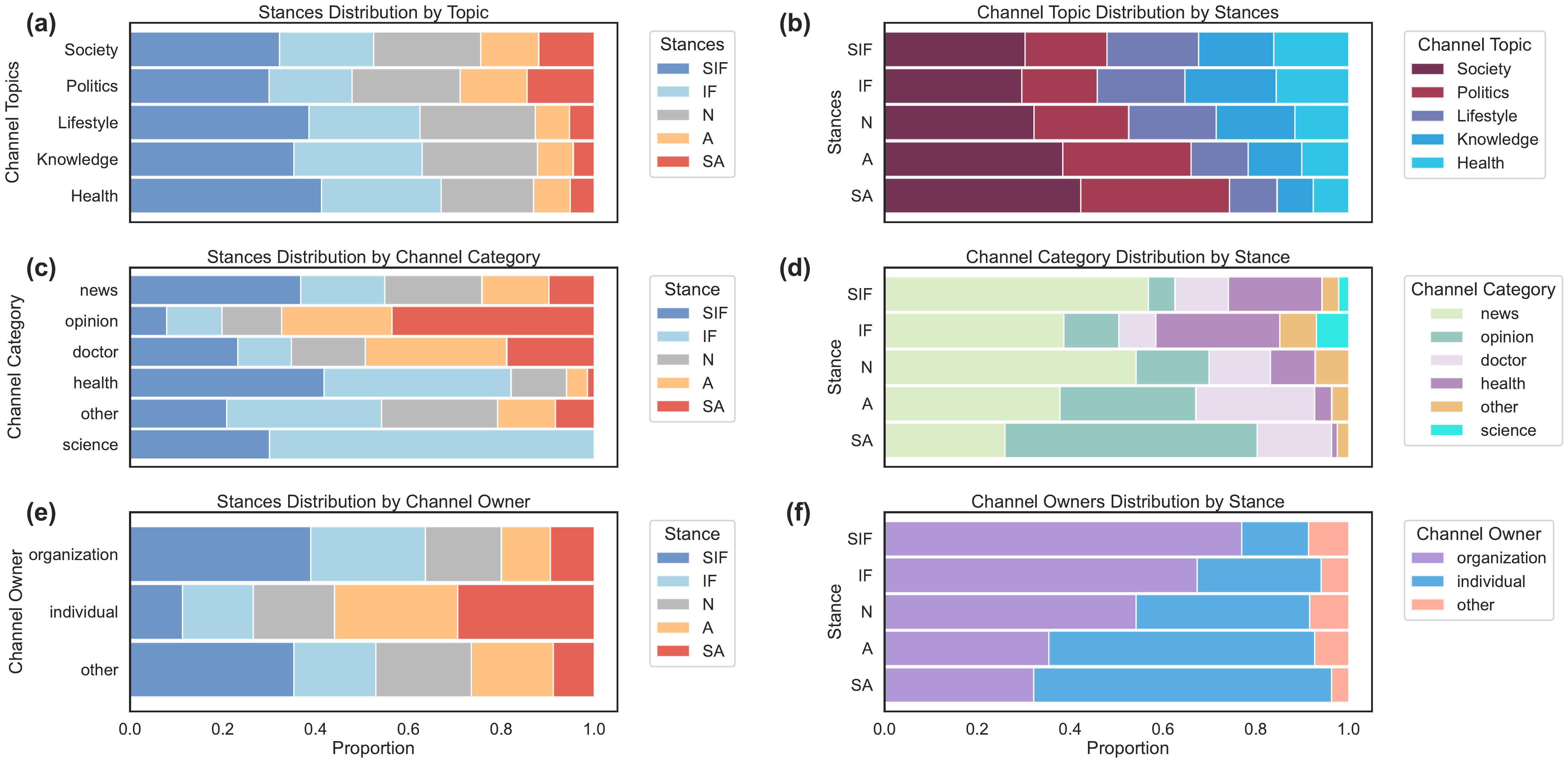}
    
    \caption{Channel characterization. Distribution of stances across different topics (a), channel categories (c), and different channel owners (e). Distribution of topics (b), channel categories (d), and channel owner types (f) across stances. The charts are row-normalized (a, c, e) and column-normalized (b, d, f). (a) shows the proportion of each stance (SIF: Strongly In Favor, IF: In Favor, N: Neutral, A: Against, SA: Strongly Against) within each topic. (b) shows the proportion of each channel owner category within each stance. (c) shows the proportion of each stance within each content category. (d) shows the proportion of each content category within each stance. (e) shows the proportion of each stance within the channel owner category. (f) shows the proportion of each channel owner category within each stance.}
    \label{fig:stance_content}
\end{figure*}

We also analyzed the distribution of stances across different channel metadata dimensions, as visualized in Figure \ref{fig:stance_content}. 
This includes channel topics (Figure \ref{fig:stance_content} a-b), manually annotated channel content categories (Figures \ref{fig:stance_content} c-d), and channel owner types (Figures \ref{fig:stance_content} e-f). 
Each pair of plots presents complementary views: the distribution of stances within each metadata category (Figures \ref{fig:stance_content} a, c, e) and the distribution of metadata categories within each stance (Figures \ref{fig:stance_content} b, d, f). 
We see a thematic heterogeneity based on stance across the channel topics. 
Videos categorized under the ``Health" topic predominantly exhibit SIF or IF stances towards vaccination. Conversely, topics related to ``Society" and ``Politics" show a higher proportion of A and SA stances. 
``Knowledge" and ``Lifestyle" topics display more intermediate distributions, generally leaning less towards hesitant stances. 
The inverse perspective, shown in Figure \ref{fig:stance_content}b, confirms this pattern: SIF and IF stances are disproportionately composed of ``Health" related content, while A and SA stances draw heavily from videos tagged with ``Society" and ``Politics".
Finally, we examined the odds ratios for the top 5 most frequent channel topics, comparing videos classified as SIF against those classified as SA, with SA serving as the reference group (Figure \ref{fig:odds_ratio_comparison}c). 
A clear thematic distinction emerged. 
Channels categorized as `Health' showed the strongest positive association with the SIF label (OR appearing visually around 2.5), followed by ``Lifestyle'' (OR $\simeq$ 2.0) and ``Knowledge" (OR $\simeq$ 1.8). 
On the other hand, channels focusing on ``Society'' (OR $\simeq$ 0.7) and ``Politics'' (OR $\simeq$ 0.5) were significantly less likely to be associated with SIF content. 
These findings highlight that channels covering societal and political discourse are disproportionately linked to content strongly opposing vaccination, whereas channels focused on health, lifestyle, and general knowledge predominantly align with pro-vaccination stances.

Analysis of manually annotated channel categories (Figure \ref{fig:stance_content}c) highlights that the ``Opinion" category is greatly associated with A and SA stances. 
In contrast, channels categorized as ``Health" and ``Science" predominantly feature SIF and IF content. 
The ``News" category displays a more mixed distribution, including substantial proportions of neutral (N), IF, and SIF stances, but also a noticeable share of A stances. 
Interestingly, the ``Doctor" category, while featuring significant favorable content, also contains a high proportion of N, A, and SA stances, suggesting creators identified as doctors are not uniformly pro-vaccination in this sample. 
Figure \ref{fig:stance_content}d reinforces these findings, showing the ``Opinion" category's dominance within A and SA stances, while ``Health", ``News", and ``Science" are the primary contributors to SIF and IF stances.

Examining the type of channel owner (Figure \ref{fig:stance_content}e), videos posted by ``Individual" creators exhibit a considerably larger proportion of A and SA stances compared to those from ``Organization" or ``Other" types. 
``Organization"-owned channels show a strong tendency towards SIF and IF stances. 
Figure \ref{fig:stance_content}f demonstrates that while ``Organization" owners contribute significantly to SIF and IF content, ``Individual" creators dominate the A and SA stance categories and also represent a substantial portion of the favorable stance video

These distributions highlight clear associations between video stance and the topic, declared content category, and owner type of the channel, suggesting systematic differences in how vaccination is framed across different types of YouTube content creators and subject areas.

\subsection{Platform Moderation}

\begin{figure}[h]
    \centering
    \includegraphics[width=0.6\linewidth]{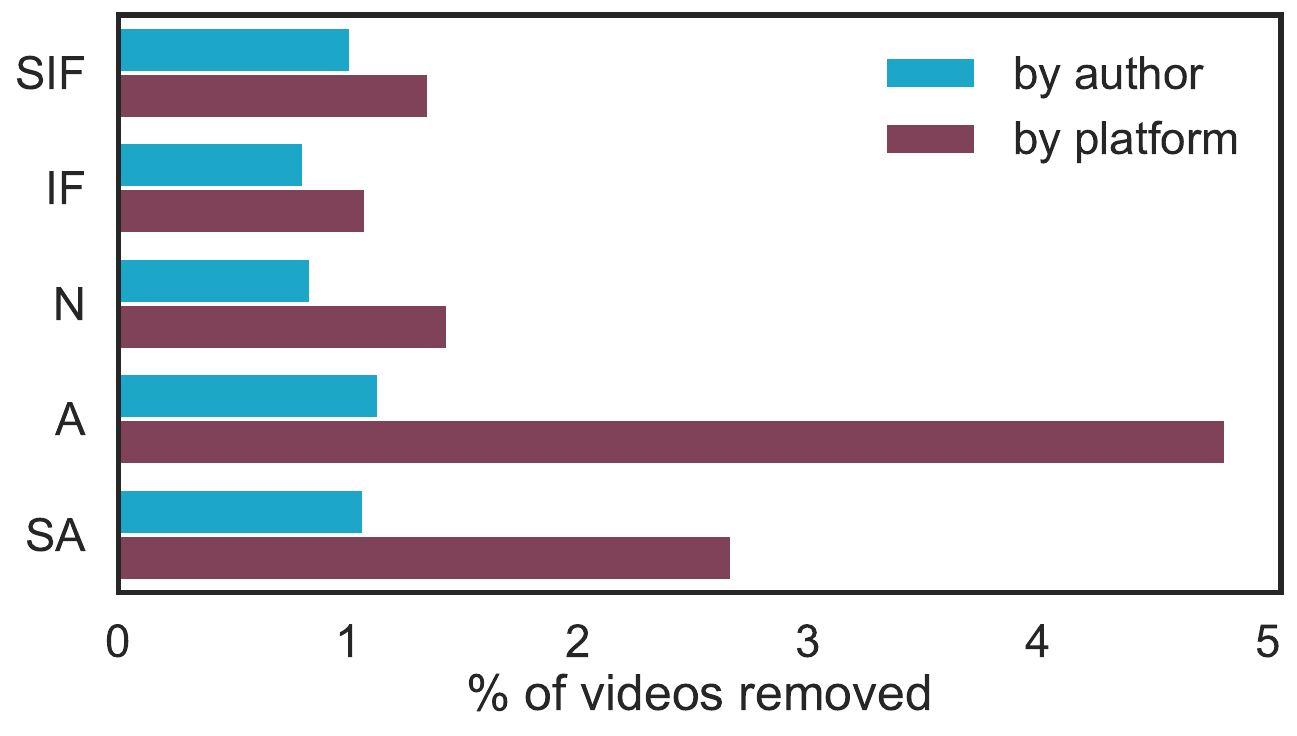}
    \caption{Percentage of videos in each category that have been removed either by the poster of the video, or by the platform.}
    \label{fig:moderation}
\end{figure}

When the metadata of the videos was re-crawled at the end of the collection, we found that 96 videos failed to return results.
We manually examined the errors the YouTube interface provided for these, and found that for 43, the platform simply tells the users that the ``This video isn't available anymore'', 24 have been made private by the uploader, 15 were unavailable due to the posting account having been terminated by the platform, 8 were removed by the uploader (a message distinct from the video simply being unavailable), 4 were removed due to violating YouTube's Community Guidelines, one because the uploader has closed their YouTube account, and one because of a copyright claim.
We consider the video disappearances attributed to the author separately from all others (which we attribute to the platform), and find rates shown in Figure \ref{fig:moderation}.
We find that, although the removals by the author of the video have about the same rate across the categories, videos labeled as A or SA were much more likely to be removed by the platform.

Considering the error codes more closely, we find that all videos removed due to violating YouTube's Community Guidelines were either labeled as A or SA. 
On the other hand, videos that belonged to channels terminated by the platform span all five categories. 
Whenever the video was made private by its author, the label is more likely to be IF or SIF (65\% of such videos) than A or SA (33\%). 
Thus, we find a variety of reasons for the disappearance of videos, with some evidence of the platform's policies affecting more the vaccine-skeptic side when it comes to sanctioning content, but not necessarily the accounts themselves.

\section{Discussion}


In this work, we analyze longitudinal data of vaccine-related content on YouTube, a platform that has become a dominant source of information and entertainment for a vast portion of the digital population. 
Our analysis shows a clear distinction in the thematic focus of the videos based on their stance on vaccination and is visually summarized in Figures \ref{fig:video_stance} and \ref{fig:stance_content}. While pro-vaccine content predominantly centers around ``Health" and specific diseases, those expressing hesitancy are more likely to fall under ``Society" and ``Politics". 
The analysis of video tags provides further evidence of the thematic divergence.
Pro-vaccine videos focus on specific diseases and health organizations, while anti-vaccine videos frequently mention political figures, slogans, and terms like ``news analysis" and ``reports". 
This finding underlines the difficulty in discerning health information from broader socio-political commentary, making it difficult for the user to identify credible sources. 
This aligns with previous results suggesting the strong politicization of health issues, particularly during COVID-19 \cite{gruzd2023facebook}. 

When analyzing the role of individual creators and organizations (Figures \ref{fig:stance_content}e-f), we discover that, perhaps unsurprisingly, established news organizations and health institutions (e.g., WHO and Forbes) are major contributors to pro-vaccine content, while individual creators with substantial followings are prominent in the ``against" and ``strongly against" content. This aligns with previous studies showing individual users ``content can be more popular even if less scientifically accurate" \cite{welbourne_science_2016}.  

An analogous picture emerges when examining the channel category (Figure \ref{fig:stance_content}e-f). 
Except for the health and science channels, which predominantly support vaccination, more heterogeneous stances characterize the remaining categories. 
The ``News" category shows a more mixed distribution, with a substantial proportion of videos in the ``neutral", ``in favor," and ``strongly in favor" categories, but also a noticeable presence of ``against" videos. 
Online news outlets might be less subject to the reliability of sources than traditional ones, reinforcing the need for users to critically evaluate the credibility of news channels.
Surprisingly, the ``doctor" category doesn't fully align with the ``health" category, falling more into the hesitant stances than the favoring ones. 
If the self-description is accurate, this could point to the personal positions of healthcare workers. 
However, the creators' self-declared certifications are difficult to verify. 
Besides, the strong association between the ``opinion" category and anti-vaccine stances indicates that a significant portion of anti-vaccine content on YouTube is not presented as factual news or scientific information, but rather as personal opinions, commentary, or analysis. 
This critical distinction suggests that much of the anti-vaccine content is not attempting to \textit{appear} as objective reporting, but it's explicitly subjective.
The prevalence of the ``Society" topic in hesitant content suggests that opposition to vaccination is often framed not as a purely scientific or medical issue, but as a broader societal concern, potentially touching on issues of individual liberty, government overreach, distrust of institutions, or cultural values. 
This framing allows anti-vaccine narratives to tap into pre-existing social and political anxieties, making them more persuasive and resistant to purely factual rebuttals. 
Videos categorized as ``Knowledge" are less likely to be hesitant. This suggests that videos focused on general knowledge dissemination, even if not directly related to health, tend to avoid promoting vaccine hesitancy. The ``Lifestyle" category is more evenly distributed, but still leans towards pro-vaccine stances. This could indicate that content creators focused on lifestyle topics (e.g., personal vlogs, daily routines) are less likely to engage in politically charged debates about vaccination, or may even subtly promote positive health behaviors.

Lastly, despite YouTube's strong policies against anti-vaccination content (especially post-COVID),\footnote{\url{https://www.bbc.com/news/technology-58743252}} our findings indicate a relatively low rate of video takedowns (2.7\%) even for content classified as ``strongly against" vaccination (20.8\% of the dataset). While we found evidence of platform intervention (videos removed for violating Community Guidelines were predominantly anti-vaccine), further labeling the data for misinformation will reveal the coverage of the platform's moderation efforts.


\textbf{Limitations}.
This study, in its three-month duration, offers only a snapshot of a rapidly evolving discourse around vaccination. Keyword-based collection might miss coded or emerging language, while LLM stance classification, despite validation, can struggle with distinguishing more complex stances. Although the scope of the study was to capture content that requires a longer user attention span, and we selected only English videos to better design the classification task, the exclusion of the YouTube Shorts format and the focus primarily one language further restricts the scope and generalizability. 

\textbf{Future work}.
Future work should prioritize longitudinal studies over longer periods to capture evolving narratives and event impacts. Investigating the interplay between video context (style, tone), stance, and user engagement metrics (views, likes, and comments) is crucial, particularly to compare the relative influence of individual creators versus organizations. Dedicated analysis of Shorts and expansion to diverse languages, regions, and platforms will also make the analysis extendable to multiple contexts.

\textbf{Ethical concerns}.
Key ethical considerations include the potential risks associated with identifying public channels, even without full anonymization, which could expose creators to unwanted attention. The research involves sensitive interpretation and framing of results, and responsible conduct requires acknowledging the broader implications of large-scale analysis of public online expression and ensuring findings are used constructively to support public health communication.

\textbf{Declaration of AI use}.
We have not used AI-assisted technologies in creating this article.


\bibliographystyle{IEEEtran}
\bibliography{referencesZotero,references}

\section*{Appendix}

\subsection*{Prompt for Stance Labeling}
\label{apx:prompt}

The prompt to ChatGPT model \texttt{gpt-4o-mini}:

\begin{quote}
{\footnotesize Your task is to determine whether the provided texts are in favor of vaccination, against vaccination, or neither.\\
For the text, assign one of the following labels based on your assessment:\\
1.``in favor of vaccination" if the overall message is that vaccines are good and helpful, and people should vaccinate. Alternatively, the text could quote or talk about somebody who is against vaccines, but if the text is against that person, the label should be``in favor of vaccination". \\
2.``against vaccination" if the overall message is that vaccines are dangerous or harmful and people should not vaccinate or given a choice not to vaccinate. Alternatively, the text could quote or talk about somebody who is in favor of vaccines, but if the text is against that person, the label should be ``against vaccination". \\
3. ``neither" if the text is not in favor or against vaccines. \\
Reply providing just the label.\\
Here is the text: \emph{[text]}}
\end{quote}

\end{document}